\documentclass[aps,prb,preprint]{revtex4-1}
\usepackage{graphicx}
\usepackage{amsmath}
\usepackage{amsfonts}
\usepackage{color}
\usepackage{mathtools}
\usepackage{braket}
\usepackage[colorlinks=true,linkcolor=blue,citecolor=blue,urlcolor=blue]{hyperref}

\begin{document}

\title{Strain-induced topological charge control in multifold fermion systems}

\author{Anumita Bose}
\affiliation{Solid State and Structural Chemistry Unit, Indian Institute of Science, Bangalore 560012, India}

\author{Awadhesh Narayan}
\email{awadhesh@iisc.ac.in}
\affiliation{Solid State and Structural Chemistry Unit, Indian Institute of Science, Bangalore 560012, India}

\date{\today}

\begin{abstract}
Multifold fermion systems feature free fermionic excitations, which have no counterparts in high-energy physics, and exhibit several unconventional properties. Using first-principles calculations, we predict that strain engineering can be used to control the distribution of topological charges in transition metal silicide candidate CoSi, hosting multifold fermions. We demonstrate that breaking the rotational symmetry of the system, by choosing a suitable strain, destroys the multifold fermions, and at the same time results in the creation of Weyl points. We introduce a low energy effective model to complement the results obtained from density functional calculations. Our findings suggest that strain-engineering is a useful approach to tune topological properties of multifold fermions.
\end{abstract} 

\maketitle

\section{Introduction} 

In recent years, an immense interest has been devoted to topological semimetals where gapless electronic phases exhibit topologically stable crossing of energy bands~\cite{armitage2018weyl,weng2016topological_1,burkov2016topological,gao2019topological}. In high energy physics, as a result of strong constraints on standard model~\cite{pal2011dirac} due to Poincar\'e symmetry, only three different types of quasi-particles, i.e., Dirac, Weyl and Majorana fermions have been predicted; though only the signature of Dirac fermions is revealed in particle physics experiments. On the other hand, in recent years great advances have been made in condensed matter physics in terms of the realisation of low energy quasi-particles, protected by less constrained crystal symmetries~\cite{neto2009electronic,lv2015experimental,huang2015weyl,hirschberger2016chiral,weng2015weyl,xu2015discovery,wan2011topological,lv2015observation,liu2014stable,liu2014discovery,wang2012dirac}. More interestingly, the crystal symmetries in solid state physics can lead to completely new quasi-particles without any high energy counterparts~\cite{bradlyn2016beyond,wieder2016double,gao2016classification} and non-zero Chern numbers, known as unconventional chiral fermions -- notable examples include, three-fold degenerate spin-1 excitations~\cite{tang2017multiple,chang2017unconventional,zhang2018double,takane2019observation,sanchez2019topological}, four-fold degenerate spin-3/2 Rarita-Schwinger-Weyl (RSW) fermions and charge-2 Dirac fermions~\cite{rarita1941theory,tang2017multiple,chang2017unconventional,pshenay2018band,takane2019observation,sanchez2019topological}, six-fold degenerate double-spin-1 excitations~\cite{tang2017multiple,bradlyn2016beyond,pshenay2018band}, predicted so far.\\ 

Though first-principle calculations predicted the presence of such unconventional chiral fermions in several space groups~\cite{bradlyn2016beyond,tang2017multiple,chang2017unconventional,zhang2018double,barman2020symmetry}, only a few materials such as AlPt~\cite{schroter2019chiral}, PdGa~\cite{sessi2020handedness}, PdBiSe~\cite{lv2019observation}, PtGa~\cite{yao2020observation},
RhSn~\cite{xu2019crystal}, CoSi~\cite{takane2019observation,rao2019observation,sanchez2019topological}, and RhSi~\cite{sanchez2019topological} have been experimentally realized, where the surface and bulk electronic properties have been visualized by scanning tunnelling microscopy and angle resolved photoemission spectroscopy. The family of transition metal silicides (MSi, M = Co, Rh) is one of the prominent proposed and experimentally verified material candidates to host unconventional chiral fermions ~\cite{tang2017multiple,chang2017unconventional,zhang2018double,changdar2020electronic,pshenay2019electronic}. The members of this family are found to host only two types of chiral fermions with opposite topological charges (without and with spin orbit coupling, Chern number, $C= \pm2$ and $\pm4$ respectively) near the Fermi energy and known to have a large topologically non-trivial energy window~\cite{yuan2019quasiparticle,sanchez2019topological} (0.85 eV for CoSi and 1.3 eV for RhSi), which provides robust quantum properties against changes in disorder and surface chemical potential. Thus, these materials satisfy the criteria to become ideal topological conductors~\cite{sanchez2019topological}.

\begin{figure*}[b]
\includegraphics[width=0.49\textwidth]{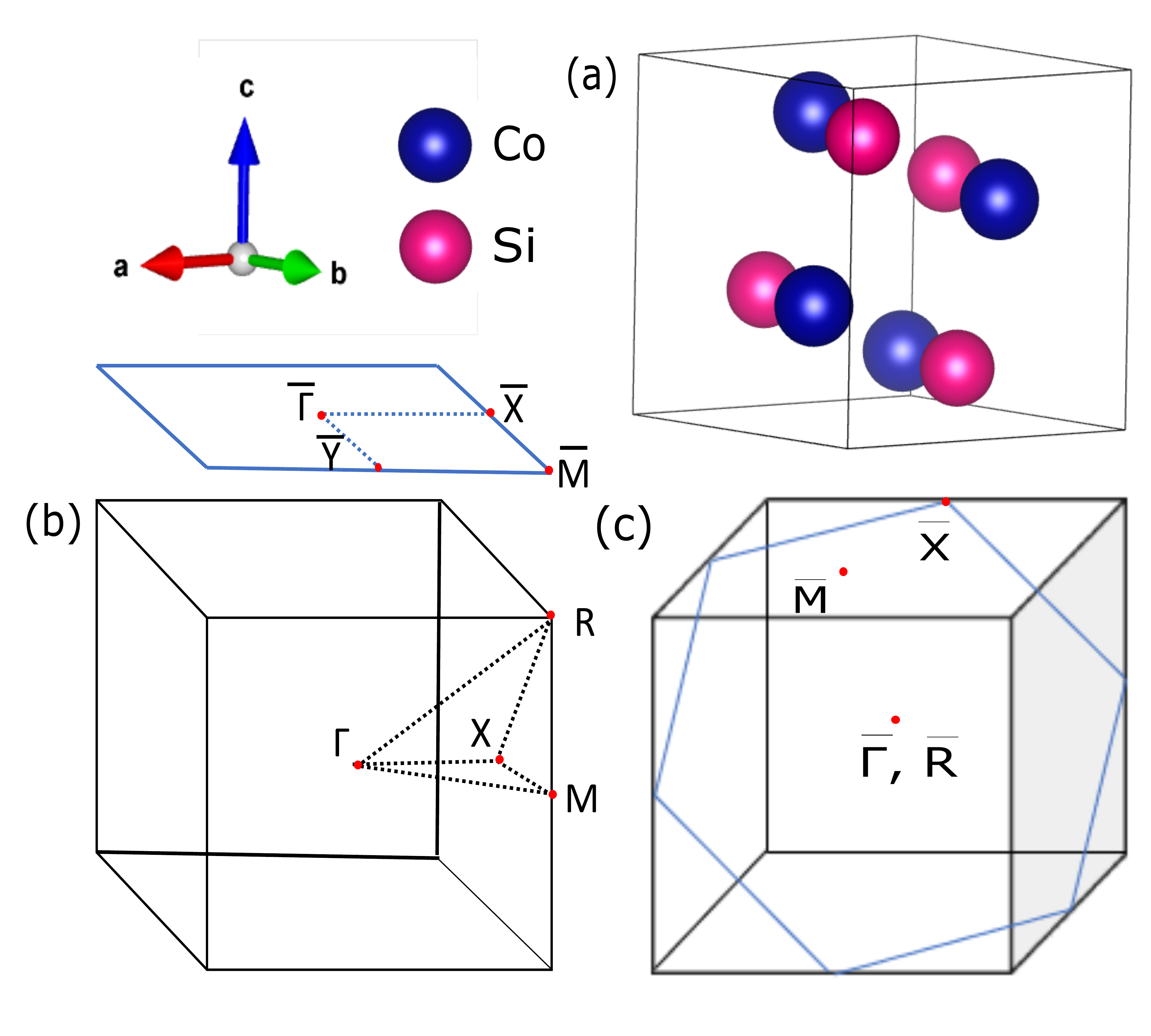}
  \caption{\textbf{Crystalline structure and Brillouin zone (BZ) of unstrained CoSi.} (a) Each simple cubic unit cell contains four Co and four Si atoms. Lattice vectors are represented as $a$, $b$ and $c$. (b) Cubic Brillouin zone with high symmetry points denoted as red dots. The projection of (001) surface is marked with blue on the top of the bulk BZ and the projected high symmetry points on this surface are labeled. (c) The hexagonal plane inside the cubic BZ marked with blue denotes the (111) plane, red dots on this surface indicate the position of high symmetry points.}\label{structure}  
\end{figure*}

\begin{figure*}
\includegraphics[scale=0.23]{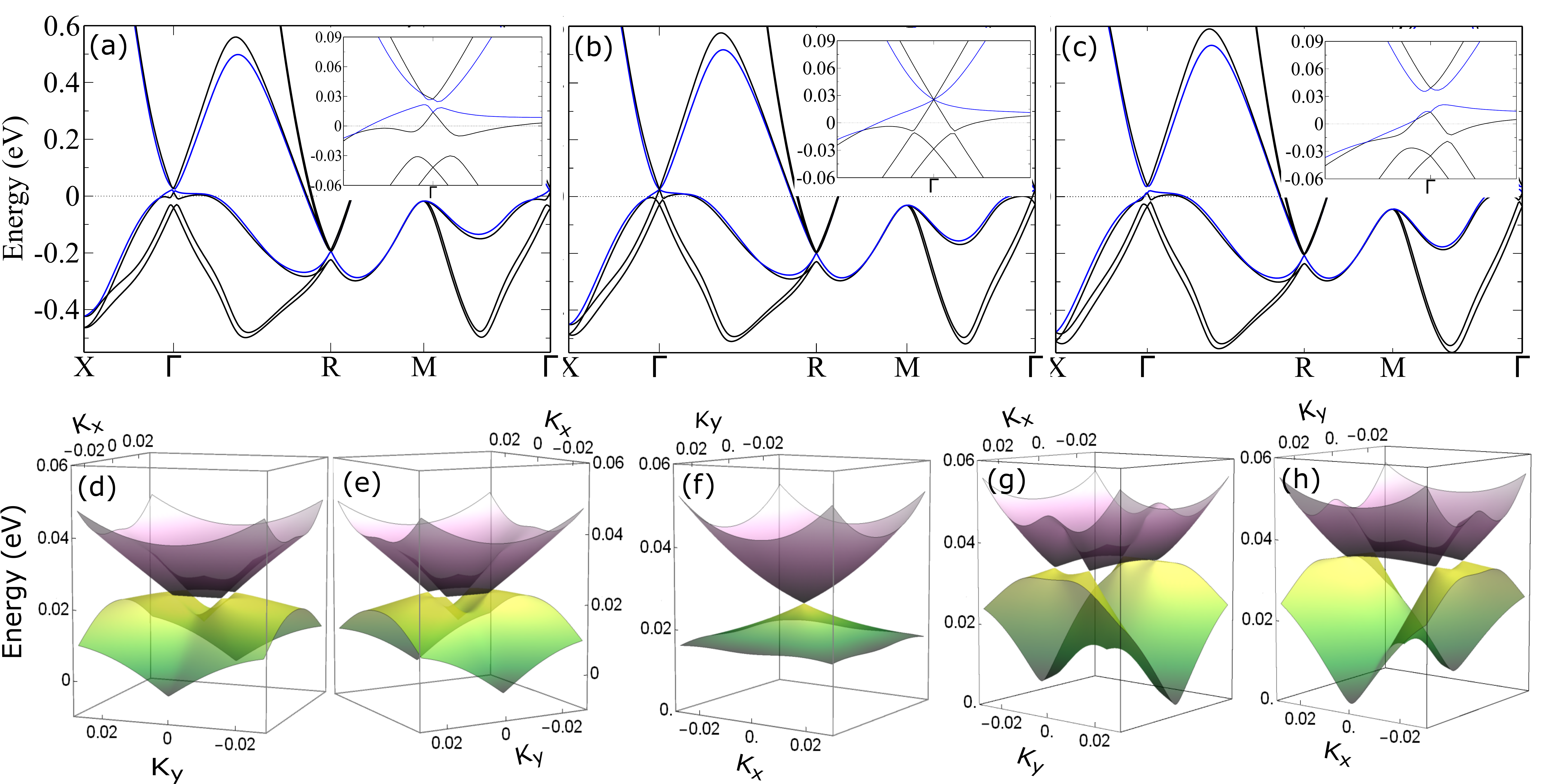}
  \caption{\textbf{Bulk electronic structure of CoSi with spin orbit coupling along the high symmetry directions.} The bulk band structures are shown for (a) 2\% tensile strained, (b) unstrained, and (c) 2\% compressive  strained systems. The figures in the insets show a zoom around at $\Gamma$ (along $X$-$\Gamma$-$R$). We obtain a four fold degeneracy at $\Gamma$ for the unstrained case as shown in (b), which is protected by a three-fold rotational and two two-fold screw symmetries along with time reversal symmetry. In contrast in (a) and (c), the degeneracies exactly at $\Gamma$  and $R$ are lifted due to the application of tensile and compressive strains, respectively, which break the three-fold rotational symmetry in the system, and a gap appears along the high symmetry direction. (d)-(h) show a three-dimensional view of the two blue colored bulk bands: (d) \& (e) for 2\% tensile strained, (f) for unstrained and (g) \& (h) for 2\% compressive strained system. For both the symmetry broken systems, we find the appearance of four two fold nodes  around $\Gamma$ between the two blue coloured bands, whereas for the unstrained systems, there is only a single node, which results in a four fold degeneracy.}
  \label{qe_bands}
\end{figure*}

These materials also feature long Fermi arcs connecting the projections of bulk nodes. In these systems, the transport properties are expected to be dominated by these chiral fermions due to the absence of any trivial surface band. These materials are not only important in theoretical studies but also provide an excellent platform for the observation of robust and unusual quantum phenomena such as quantized circular photogalvanic effect~\cite{ni2020giant,le2020ab}, high optical conductivity~\cite{li2019optical,ni2020linear,maulana2020optical} and large unsaturated magnetoresistance~\cite{xu2019quantum}. Recent studies on the materials of this family also include theoretical investigation of the dynamical conductivity as a function of photon energy ~\cite{habe2019dynamical}, their Floquet engineering~\cite{jaiswal2020floquet}, and observation of quantum oscillations in thermoelectric signals due to high carrier mobility in CoSi~\cite{xu2019crystal}, as well as control over maximal Chern number tuned by handedness of the crystal structure in PdGa~\cite{schroter2020observation,sessi2020handedness}. Furthermore, six-fold fermions with no protected Fermi arcs in achiral PdSb$_{2}$ ~\cite{sun2020direct} and a prediction of single linear dispersive six-fold excitation with trivial surface Fermi arcs but non-trivial hinge arcs in electrides~\cite{nie2020six} appear to be interesting recent studies on multifold fermions protected by non-symmorphic crystal symmetries. Apart from the fermions with non-zero topological charge, several materials including WC~\cite{ma2018three}, MoP~\cite{lv2017observation}, $\theta$-TaN~\cite{weng2016topological}, ZrTe~\cite{weng2016coexistence}, YRh$_{6}$Ge$_{4}$~\cite{zhu2020evidence}, with symmorphic crystal symmetry are known to host \textit{new fermions} with triply degenerate band crossing points. These materials host protected surface Fermi arcs and gapless Landau levels in presence of symmetry preserving magnetic fields, suggesting the possibility of observing transport anomalies. Extremely high mobility and conductivity have been observed in these materials~\cite{kumar2019extremely}, and a doping-driven Lifshitz transition is also predicted in these class of materials~\cite{zhu2016triple}. \\

Not only the presence of symmetries, but also their breaking can result in new phases of matter, and cause topological phase transitions. Application of mechanical strain~\cite{shao2017strain,winterfeld2013strain,nie2020six}, chemical strain~\cite{narayan2014topological,hsieh2012topological,dziawa2012topological}, external pressure~\cite{bahramy2012emergence,zhang2015breakdown,he2015pressure}, electric~\cite{collins2018electric,kim2012topological} or magnetic~\cite{zhang2019possible} fields are found to be such symmetry breaking parameters to lower the symmetries of a system and yield different topological phases. Strain switching can act as a useful tool to engineer transitions between different topological phases in several systems, such as methyl-decorated SiGe films~\cite{teshome2018strain}, bilayers of group IV and V elements~\cite{huang2014strain}, bismuth–tellurohalide–graphene
heterostructures~\cite{tajkov2019uniaxial}, $\beta$-As$_{2}$Te$_{3}$~\cite{pal2014strain},
ZrTe$_5$~\cite{mutch2019evidence}. Besides, different types of nodal lines~\cite{wang2020strain} can be realized in quasi 2D $\alpha$-FeSi$_2$ as a result of strain control. Tunning of type-II Dirac point in NiTe$_2$ ~\cite{ferreira2021strain} and surface Dirac point~\cite{zeljkovic2015strain} in topological crystalline insulators are also achievable by controlling external strain.
In monolayer black phosphorus, mechanical strain causes Majorana zero energy edge modes~\cite{alidoust2018strain} and also changes in the symmetry of superconducting pairing from s and p-wave to d and f-wave, respectively~\cite{alidoust2019control}.\\

Motivated by these strain-controlled rich phenomena, in this paper, we consider symmetry breaking in transition metal silicide, CoSi, by applying external strain. We apply biaxial strain (both tensile and compressive upto 2\%) in the $xy$ plane to break rotational symmetry along the 111 axis, at the same time preserving the screw symmetries. From our first-principles calculations, we find that lowering the symmetry by the application of external strain not only causes a redistribution of the Chern number, but this allows control over the distribution of these Weyl points in the momentum space. We also investigate the characteristic feature of unconventional fermions, i.e., long Fermi arcs which are accompanied by trivial closed Fermi arcs and their evolution in strained systems. We use a low energy effective model augmented by symmetry breaking terms to understand our \textit{ab initio} results and complement it by a symmetry analysis. Our results show that application of strain is a promising approach to tune topological properties of multifold fermions. 

\begin{figure*}[ht]
\includegraphics[width=0.48\textwidth]{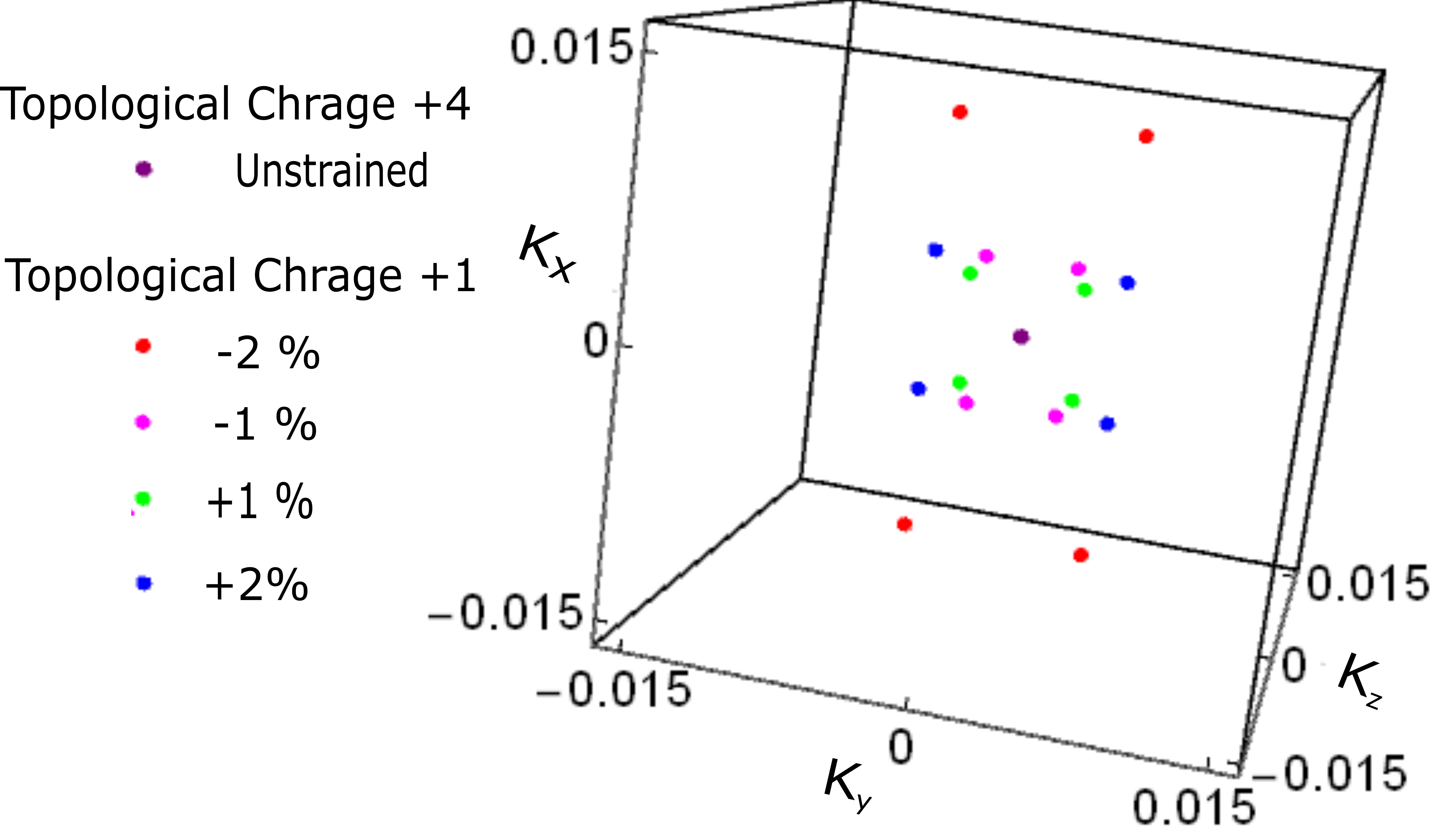}
  \caption{\textbf{Distribution of Weyl points near $\Gamma$} Red, magenta, green and blue dots denote the location of Weyl points with topological charge +1 around the $\Gamma$ point in the 3D momentum space between the two blue colored bands for -2\%, -1\%, 1\% and 2\% strained systems, respectively. The purple dot at the center denotes the $\Gamma$ point having four fold degenerate spin-3/2 RSW fermion with topological charge +4 in the unstrained system. For the strained systems, due to the breaking of three fold rotational symmetry, four fold degeneracy is destroyed and four spin-1/2 Weyl points with topological charge +1 distributed symmetrically around the BZ center, are obtained. The distribution of these Weyl points are controlled by the strain.}\label{nodes_Gamma}  
\end{figure*}

\begin{figure*}
\includegraphics[scale=0.25]{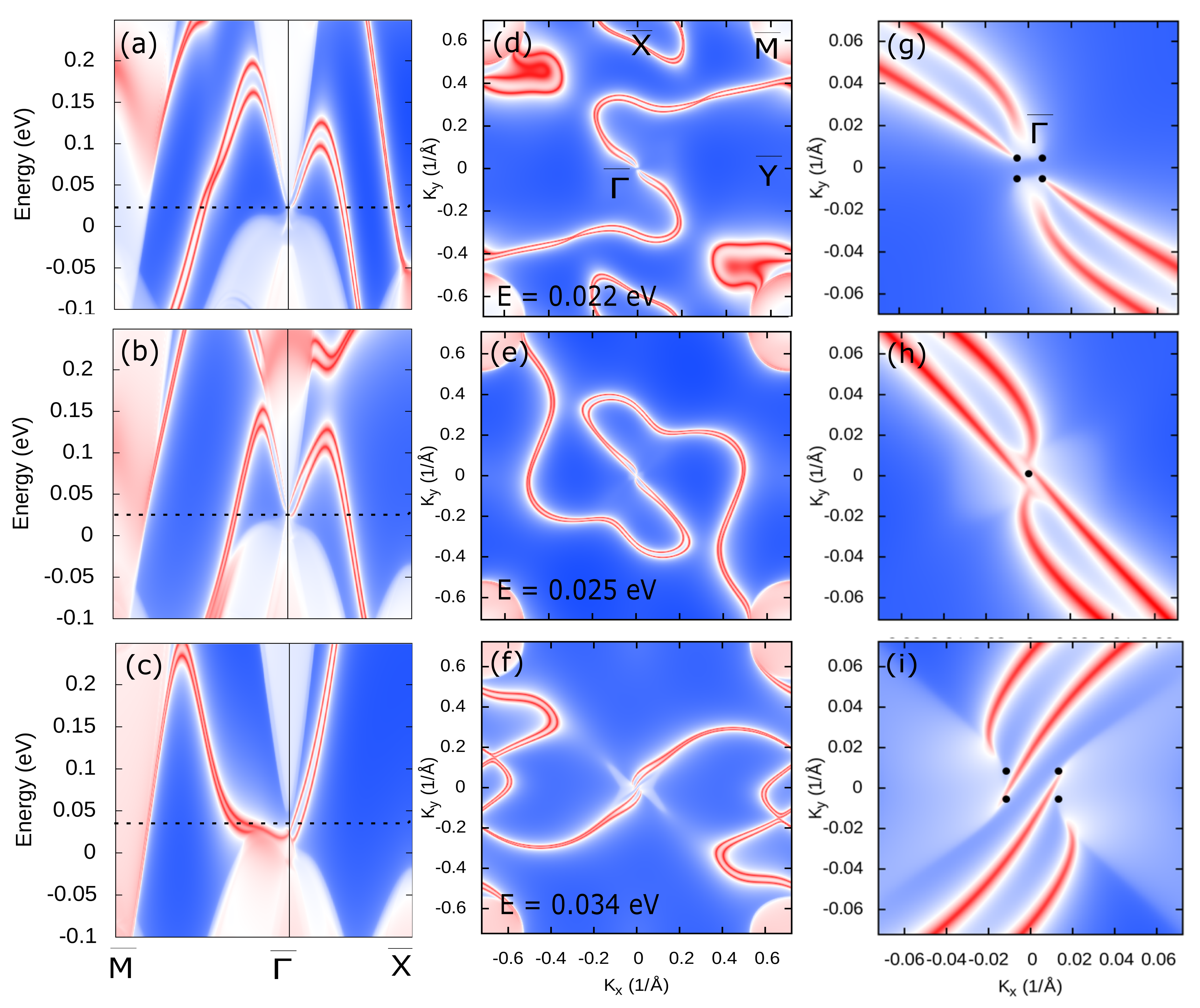}
  \caption{\textbf{Surface spectra and Fermi arcs on the (001) surface.} (a)-(c) Surface states on (001) surface for the semi-infinite 2 \% tensile strained, unstrained and 2 \% compressive strained system respectively along $\Bar{M}-\Bar{\Gamma}-\Bar{X}$ direction. Dashed lines in the left panel denote the energy cut at which we performed the Fermi arc calculation. (d)-(f) Four non trivial arc states connecting the bulk projections $\Bar{\Gamma}$ and $\Bar{M}$ at the energy of the nodes for 2\% tensile strained, unstrained, 2\% compressive strained systems. Panels (g)-(i) are the respective zoom around the BZ center of the (001) surface of these three systems. For the unstrained system [(e)], four arcs connect topological charges +4 (at $\Gamma$) and -4 (at $R$). For the strained systems, four non-trivial arcs connect the four pairs of Weyl nodes distributed around $\Gamma$ and $R$ point. Black dots in (g)-(i) show the nodes in the lower panel of Fig.\,\ref{qe_bands}.} 
  \label{fermi_arcs}
\end{figure*}

\section{Methods} 

We have carried out the first-principles calculations in the density functional theory framework encoded in the Quantum Espresso package~\cite{giannozzi2009quantum,giannozzi2017advanced}, using the Perdew-Burke-Ernzerhof exchange-correlation functional~\cite{perdew1996generalized} and ultra-soft pseudopotentials~\cite{vanderbilt1990soft}. For the bulk calculations a plane wave cutoff of $60$ Ry was chosen and an $8\times8\times8$ $\Gamma$-centered $k$-point grid was used for self-consistent calculations. During structural optimization of the strained systems, atomic coordinates were allowed to relax until forces became less than $10^{-3}$ Ry/bohr. We constructed Wannier functions based tight binding models obtained from maximally localized Wannier functions using the {\sc wannier90} code~\cite{mostofi2008wannier90} with Co 3$d$ and Si 3$p$ orbitals as the basis.

Positions of the nodes were obtained by using the Nelder and Mead’s Downhill Simplex Method~\cite{nelder1965mead} implemented in WannierTools~\cite{wu2018wanniertools}, taking a uniform mesh in the three-dimensional BZ as a set of initial points. After finding the nodes, a small sphere is constructed around each of the nodes. An evolution of the sum of hybrid Wannier charge centers as a function of the polar angle is then tracked to evaluate the enclosed Weyl point's chirality. Surface spectrum was obtained by calculating the surface Green's function for semi-infinite systems using iterative Green's function method~\cite{guinea1983effective,sancho1984quick,sancho1985highly}, implemented in WannierTools package. \\

\section{Results and Discussion} 

%symmetry of the system
\subsection{Crystal structure and symmetries}

CoSi crystallizes in the cubic structure with lattice constant, $a$ = 4.43 \text{\AA}, and space group $P2_13$ (number 198). At the Brillouin zone (BZ) center, $\Gamma$, this space group has three generators: two fold non-symmorphic screw rotations along $z$ and $y$ axes: $S_{2z} = \{C_{2z}| \frac{1}{2}, 0, \frac{1}{2}\}$ and $S_{2y} = \{C_{2y} | 0, \frac{1}{2}, \frac{1}{2}\}$; and a three fold rotation: $S_3 = \{C_3 | 0,0,0\}$. On the other hand, the generators at the corner of the BZ, i.e., at the $R$ point are: $S_{2x} = \{C_{2x} | \frac{1}{2}, \frac{3}{2}, 0\}$, $S_{2y} = \{C_{2y} | 0, \frac{3}{2}, \frac{1}{2}\}$ and $S_3 = \{C_3^{-1} | 0,1,0\}$. However at $M$, there is no three-fold rotational symmetry and this high-symmetry point only holds non-symmorphic screw symmetries: $S_{2z} = \{C_{2z} | \frac{1}{2}, 0, \frac{1}{2}\}$ and $S_{2y} = \{C_{2y} | 0, \frac{1}{2}, \frac{1}{2}\}$. \\

In our work, a biaxial strain is applied in the $xy$ plane and the system is relaxed along the $z$ direction (Lattice parameters and atomic positions for both the strained \& unstrained systems are given in TABLE \ref{table:unitcell} and \ref{table:atomic_coordinates} in the Appendix). For such a strain configuration the system loses the cubic symmetry. For 2\% tensile strained system, value of the in-plane lattice parameters, $a$ and $b$, are increased and become 4.52 \text{\AA}, whereas the lattice parameter along the perpendicular direction, $c$, decreases and takes the value 4.36 \text{\AA}. The situation gets reversed for 2\% compressive strained system, where $a$ and $b$ are 4.34 \text{\AA} and $c$ becomes 4.49 \text{\AA}  after relaxation. Application of either compressive or tensile strain breaks the three fold rotational symmetry of the system keeping the two fold screw symmetries similar to that of space group $P2_{1}2_{1}2_{1}$ (number 19). Further at these high symmetry points time reversal symmetry (TRS) continues to be obeyed.
\label{structure_and_symmetry}

%band structures
\subsection{Bulk electronic structure} 

We begin our analysis by examining the bulk electronic structure of CoSi. Our calculated band structure with spin orbit coupling (SOC) is presented in Fig.\,\ref{qe_bands}. Around the Fermi level, the contributions to the electronic states come from a hole pocket at the $\Gamma$ point and an electron pocket at the $R$ point. For the unstrained system, four fold degeneracy at $\Gamma$ and six fold degeneracy at $R$ point lead to an unconventional spin-3/2 RSW fermion and a double spin-1 fermion, respectively, as shown in Fig.\,\ref{qe_bands}(b). The band structure along the high symmetry directions for 2\% tensile and compressive strained systems are shown in Fig.\,\ref{qe_bands}(a) and (c), respectively. In both these cases, the characteristic four fold degeneracy at $\Gamma$ point and six fold degeneracy at $R$ point are lost as a result of biaxial strain. Instead we obtain a gap along the high symmetry direction. Although along other directions, there exist multiple two fold degeneracies around these time reversal invariant momenta (TRIM). Three-dimensional views of the two blue coloured bands around $\Gamma$ of Fig.\,\ref{qe_bands}(a)-(c), are shown in Fig.\,\ref{qe_bands}(d)-(h). For both the strained systems [Fig.\,\ref{qe_bands}(d)-(e) for tensile strained and Fig.\,\ref{qe_bands}(g)-(h) for compressive strained], there are four nodes between these two bands, all of which turn out to carry a topological charge +1. The nodes for the 2\% tensile strained and 2\% compressive strained systems are obtained at the energies of 0.022 eV and 0.034 eV, respectively. The position of four-fold degeneracy at $\Gamma$ in the unstrained system is 0.025 eV. We will later present the calculated Fermi arcs of these systems at these particular energy values.

Fig.\,\ref{nodes_Gamma} shows the distribution of these nodes around $\Gamma$ in the three dimensional momentum space. The purple dot at the center of the three dimensional momentum space, i.e., the $\Gamma$ point, denotes the four fold degenerate spin-3/2 degeneracy point for the unstrained system. For the strained systems Weyl nodes are distributed in the $k_{x}-k_{y}$ plane in a systematic way. Red, magenta, green and blue dots denote  the Weyl points near $\Gamma$ for the -2, -1, 1 and 2\% strained systems respectively. With increasing strain magnitude, the nodes move away from the $\Gamma$ point. \\ 
We also obtained four spin-1/2 Weyl nodes with chirality -1 around $R$ point between those two selected bands. However, since $R$ point is situated at the corner of the BZ, four spin-1/2 Weyl nodes were observed near four different corners of the selected BZ, e.g. in case of 2\% tensile strained system, the nodes are located at: (-0.499, 0.499, 0.497), (-0.499, -0.499, -0.497), (0.499, 0.497, 0.499) and (-0.499, 0.499, -0.497).
Though all the higher order degeneracies at the high symmetry points are broken as a result of three fold symmetry breaking, the band degeneracy at M is still present, since this degeneracy is protected only by screw symmetries, which are preserved in the strained systems too. \\

\begin{figure*}
\includegraphics[scale=0.2]{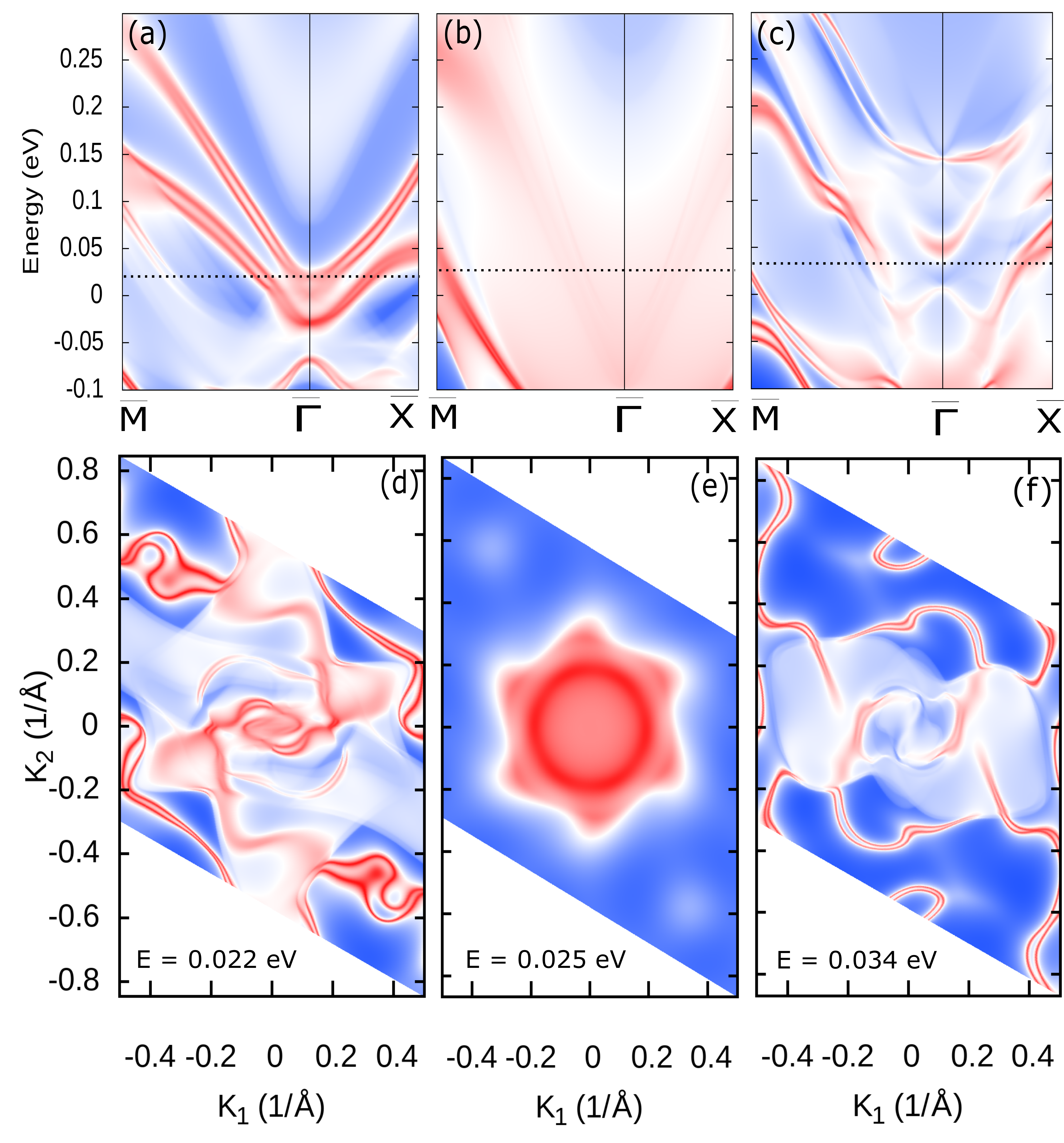}
  \caption{\textbf{Surface states and Fermi arcs on the (111) surface.} (a)-(c) represent the surface bands for the semi-infinite geometry. Dashed lines in (a)-(c) denote the energy cut at which we performed the Fermi arc calculation. Fermi arcs at constant energy are depicted in the lower panel. (d) 2\% tensile strained system at energy 0.022 eV, (e) unstrained system at 0.025 eV and (f) 2\% compressive strained system at energy 0.034 eV. $K_1$ and $K_2$ denote the in-plane momenta on this surface. Bulk points $\Gamma$ and $R$ being projected on the same point, only trivial closed loop appears for the unstrained system as shown in (e), as these two points carry topological charge +4 and -4 respectively. In case of strained systems, nodes with topological charge $\pm1$ are distributed around the high symmetry points $\Gamma$ and $R$, giving rise to small arcs near the center of the surface BZ.} 
  \label{fermi_arcs_111}
\end{figure*}

\subsection{Nature of surface states}

In order to carry out a topological analysis of the electronic structure obtained from DFT, we studied the surface properties for the semi-infinite system. Fig.\,\ref{fermi_arcs} shows surface excitations for both the unstrained and strained systems. We first study the surface properties for the (001) surface, where the projection of bulk $\Gamma$ and $R$ points are $\Bar{\Gamma}$ and $\Bar{M}$ respectively [as shown in Fig.\,\ref{structure}(b)]. We expect surface Fermi arcs to connect non-trivial topological nodes with opposite chirality.  In Fig.\,\ref{fermi_arcs} (a)-(c), we present surface spectra for semi-infinite 2\% tensile, unstrained and 2\% compressive strained CoSi respectively on the (001) surface. Surface states as a function of in-plane momenta at the energy of bulk nodes are shown in Fig.\,\ref{fermi_arcs}(d)-(f). Fig.\,\ref{fermi_arcs}(g)-(i) depict the zoomed-in view of Fig.\,\ref{fermi_arcs}(d)-(f), respectively, around the $\Gamma$ point. The black dots in these three figures denote the projection of nodes on the (001) surface. We find four topologically non-trivial extended Fermi arcs from  $\Bar{\Gamma}$ to $\Bar{M}$ connecting bulk nodes $\Gamma$ with topological charge +4 and $R$ with topological charge -4 in the unstrained CoSi\cite{tang2017multiple, pshenay2018band}, shown in Fig.~\ref{fermi_arcs}(e) and (h). At the scale of full BZ, the change in the Fermi arcs in the strained systems is quite small due to their extended nature. In the two strained cases (tensile and compressive) the four non trivial arcs connect the Weyl points distributed around $\Gamma$ and $R$ with opposite chirality. Additionally, unlike the unstrained system, two trivial Fermi arcs appear along the $\Bar{X}$-$\Bar{M}$ direction [Fig.~\ref{fermi_arcs}(d)] for 2\% tensile strained system and along $\Bar{Y}$-$\Bar{M}$ direction [Fig.~\ref{fermi_arcs}(f)] in 2\% compressive strained systems.\\

In order to establish the occurrence of Weyl node distribution, we studied the surface states on (111) surface as well. Fig.\,\ref{fermi_arcs_111}(a)-(c) show the surface bands for the 2\% tensile trained, unstrained and 2\% compressive strained systems, respectively. (d)-(f) show the respective Fermi arcs. On the (111) surface, the two bulk high symmetry points $\Gamma$ and $R$ project to the center of the surface BZ [denoted as $\Bar{\Gamma}$ or $\Bar{R}$ in Fig.~\ref{structure}(c)]. For the unstrained system, $\Gamma$ and $R$ carry topological charge +4 and -4 respectively. So, at the $\Bar{\Gamma}$ (or $\Bar{R}$), the topological charges essentially overlap with each other. As a result, unlike (001) surface, here we obtain trivial closed Fermi arcs~\cite{yuan2019quasiparticle}, which are shown in Fig.~\ref{fermi_arcs_111}(e). However, in the case of strained systems, nodes being situated away from the high symmetry points, $\Gamma$ and $R$, we expect small Fermi arcs connecting Weyl points with opposite chirality near the center of the surface BZ. Indeed, this is the case as depicted in Fig.\,\ref{fermi_arcs_111}(d) \& (f), where we observe small arcs around the BZ center, unlike the closed loop in (e) for unstrained system. 

To support our findings in semi infinite slabs, we also calculated the band structures for finite sized systems, which are shown in Fig.\,\ref{surface_001_111} in the Appendix.
\label{surface}\\

\begin{figure*}
\includegraphics[scale=0.25]{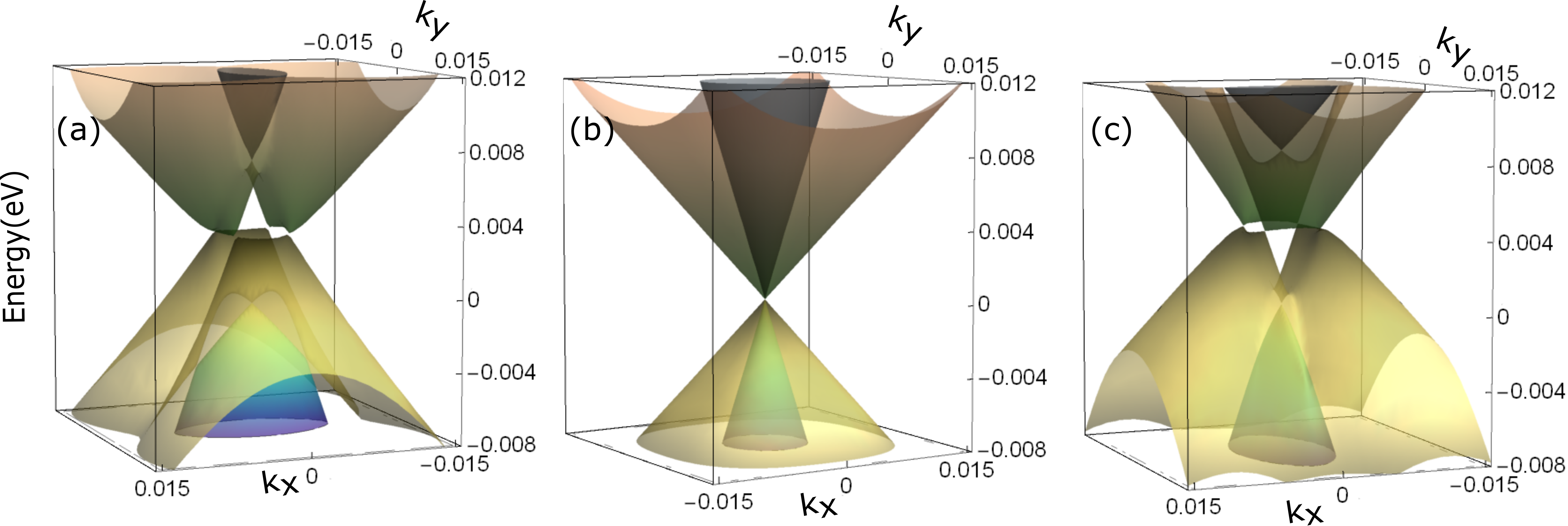}
  \caption{\textbf{Band structure from the low-energy effective model.} (b) Band structure showing a four fold degeneracy at $\Gamma$ for unstrained CoSi, where two fold screw symmetries as well as three fold rotational symmetry are protected. (a) \& (c) show band structure for the 2\% tensile and compressive strained system with broken rotational symmetry, showing four two fold nodes between the two middle bands around the center of the BZ. Due to the application of external strain, the four fold degeneracy in (b) is destroyed and in turn four two fold nodes are created around $\Gamma$ in (a) and (c).}
  \label{model}
\end{figure*}

%model Hamiltonian
\subsection{Model Hamiltonian and symmetry arguments} 

In order to have a deeper understanding of the topology of the strained systems in comparison to the unstrained one, we use a low energy model Hamiltonian around the center of the Brillouin zone ($\Gamma$). First, for the unstrained system we consider the following low energy linear $\textbf{k.p}$ model Hamiltonian~\cite{pshenay2018band}

\begin{widetext}
\begin{eqnarray}
H_0 =
\begin{pmatrix}
	ak_{z} & a(k_{x}-ik_{y}) & b(z_{1}k_{x}-z_{2}k_{y}) & bk_{z} \\ 
	a(k_{x}+ik_{y}) & -ak_{z} & bk_{z} & -b(z_{1}k_{x}+z_{2}k_{y}) \\ 
	b^*(z^*_{1}k_{x}-z^{*}_{2}k_{y}) & b^*k_{z} & -ak_{z} & -a(k_{x}+ik_{y}) \\
	b^*k_{z} &  -b^*(z^{*}_{1}k_{x}+z^{*}_{2}k_{y}) & -a(k_{x}-ik_{y}) & ak_{z}
\end{pmatrix},
\label{eq:2}    
\end{eqnarray}
\end{widetext}

where $a$ is a real number whereas $b$, $z_{1} = e^\frac{i\pi}{3}$ and $z_{2} = e^\frac{i\pi}{6}$ are complex. We briefly summarize the rationale behind the derivation of the $H_0$. We used the conjugated irreducible representations $\Bar{\Gamma_6}$ and $\Bar{\Gamma_7}$ presented in Bilbao crystallographic server~\cite{aroyo2011crystallography,elcoro2017double}. The matrix representing the four-fold degeneracy at $\Gamma$ in equation~\eqref{eq:2} respects two fold screw symmetries: $S_{2z}= -i(\sigma_{z} \otimes \sigma_{x})$,
$S_{2y}= -i(\sigma_{0} \otimes \sigma_{y})$ and $S_{2x}= -i(\sigma_{z} \otimes \sigma_{z})$  and the three fold rotational symmetry, $S_3$,

\begin{align}
\centering
S_3 = \frac{1}{\sqrt{2}}
\begin{pmatrix}
	e^\frac{-i5\pi}{12} & e^\frac{i\pi}{12} \\ 
	e^\frac{-i5\pi}{12} & e^\frac{-i11\pi}{12} 
\end{pmatrix}
\oplus \frac{1}{\sqrt{2}}
\begin{pmatrix}
	e^\frac{i5\pi}{12} & e^\frac{-i\pi}{12} \\ 
	e^\frac{i5\pi}{12} & e^\frac{i11\pi}{12} 
\end{pmatrix},
\label{eq:1} 
\end{align}

The low energy model Hamiltonian, constrained by the symmetry operator $S$, should satisfy $SH(k)S^{-1}=H(Sk)$. Now the under the action of $S_{2x}$, $k_{x} \longrightarrow k_{x}$, $k_{y} \longrightarrow -k_{y}$ and $k_{z} \longrightarrow -k_{z}$. So, the coefficients of $k_x$ should commute with $S_{2x}$. Again, for $S_{2y}$, $k_{x} \longrightarrow -k_{x}$, $k_{y} \longrightarrow k_{y}$ and $k_{z} \longrightarrow -k_{z}$. This indicates that the coefficients of $k_x$ should anticommute with $S_{2y}$. The allowed coefficients of $k_x$, obeying these constraints are: $(\sigma_{0} \otimes \sigma_{x})$, $(\sigma_{z} \otimes \sigma_{x})$, $(\sigma_{x} \otimes \sigma_{z})$ and $(\sigma_{y} \otimes \sigma_{z})$.\\

Eigenenergies of $H_0$ form a four fold degeneracy at $\mathbf{k}=0$, as shown in Fig.\,\ref{model}(a) (parameters listed in TABLE \ref{table:parameters}.).

As discussed in section \ref{structure_and_symmetry}, due to application of biaxial strain, the three fold rotational symmetry is broken, although the screw symmetries are still protected. Our  Hamiltonian should be such that it commutes with the screw operators but doesn't commute with the three-fold rotational operator, $S_3$. In order to incorporate the effect of strain, we use the strain tensor introduced in Ref.\cite{nikolaev2021effect}, which acts as a perturbation on the unstrained Hamiltonian. Biaxial strain in the $xy$ plane can, in effect, be modelled by a uniaxial strain along the $z$ direction. Our perturbative term reads,

%\begin{widetext}
%\begin{eqnarray}
%H_1 =
%\begin{pmatrix}
    %d(\epsilon_{xx}+\epsilon_{yy}) & 0 & 0 & c(\epsilon_{xx}+z_{1}\epsilon_{yy}) \\ 
	%0 & d(\epsilon_{xx}+\epsilon_{yy}) & -c(\epsilon_{xx}+z_{1}\epsilon_{yy}) & 0 \\ 
	%0 & -c^{*}(\epsilon_{xx}+z^{*}_{1}\epsilon_{yy}) & d(\epsilon_{xx}+\epsilon_{yy}) & 0 \\
	%c^{*}(\epsilon_{xx}+z^{*}_{1}\epsilon_{yy}) &  0 & 0 & d(\epsilon_{xx}+\epsilon_{yy})
%\end{pmatrix},
%\label{eq:3}    
%\end{eqnarray}
%\end{widetext}

\begin{widetext}
\begin{eqnarray}
H_1 =
\begin{pmatrix}
    d\epsilon_{zz} & 0 & 0 & cz_{1}^{2}\epsilon_{zz} \\ 
	0 & d\epsilon_{zz} & -cz_{1}^{2}\epsilon_{zz} & 0 \\ 
	0 & -c^{*}{z^{*}_{1}}^{2}\epsilon_{zz} & d\epsilon_{zz} & 0 \\
c^{*}{z^{*}_{1}}^{2}\epsilon_{zz} &  0 & 0 & d\epsilon_{zz}
\end{pmatrix},
\label{eq:3}    
\end{eqnarray}
\end{widetext}

where $c=c_{0}e^\frac{-i\pi}{6}$ is complex and $d$ is real parameter. Here $\epsilon_{zz}$ denotes the effective strain along the $z$ direction, which can be obtained from the lattice parameters of the system. The total Hamiltonian for the strained system is expressed as: $H= H_{0} + H_{1}$. The energy eigenvalues of the perturbed Hamiltonian H are shown in Fig. \,\ref{model}  (a) and (c). We used the parameters as listed in TABLE \ref{table:parameters}.

%\begin{widetext}
%\begin{table}[h]
%\begin{tabular}{|p{1.8cm}|p{0.9cm}|p{0.9cm}|p{0.9cm}|p{1.0cm}|p{0.9cm}|}
 
 \begin{table*}[h]
 \caption{Values of parameters for the model Hamiltonian.}
\label{table:parameters}
\begin{tabular}{|p{2.8cm}|p{1.8cm}|p{1.8cm}|p{1.8cm}|p{1.8cm}|p{1.8cm}|}

 \hline
 Strain values & $a$ (eV) & $b$ (eV)& $\epsilon_{zz}$ &$c_0$(eV)& $d$(eV)\\
 \hline
  $+2\%$  &0.56&1.19& -0.017&0.235&-0.182\\
 
  Unstrained &0.56& 1.19& 0&0&0\\

  $-2\%$ &0.56& 1.19& 0.014&0.3&0.31\\

 \hline
\end{tabular}
\end{table*}
%\end{widetext}

After having constructed the low energy model and understood the nature of the band structure based on it, we next consider general symmetry arguments to gain further insights.
For transition metal silicides, without considering the SOC, at $\Gamma$, the two fold screw operators ($S_{2y}$ and $S_{2z}$) commute, which indicates the presence of a simultaneous eigenstate, say, $\ket{\psi}$. Further, the relations between $S_3$, $S_{2y}$ and $S_{2z}$ ensure the existence of other two mutual eigenstates $\ket{S_3\psi}$ and $\ket{S_3^2\psi}$ of the two screw operators~\cite{barman2020symmetry}. The non triviality of both these screw symmetries and the fact that $S_3$ commutes with the Hamiltonian, confirm the degeneracy of these three distinct eigenstates $(\ket{\psi}, \ket{S_3\psi}, \ket{S_3^2\psi})$ at this point. This three-fold degeneracy is essentially the spin-1 fermion in the unstrained system without SOC~\cite{tang2017multiple, barman2020symmetry}. Now, upon applying strain in the system, $S_3$ no longer commutes with the Hamiltonian. So, the degeneracy between these three mutual eigenstates of the screw operators is lifted at $\Gamma$. This was the consideration without SOC. Next, in the spin orbit coupled system, TRS satisfies, $\tau^{2} = -1$. Therefore, except at the TRIM points, Kramer's degeneracy is lifted throughout the BZ. Thus, TRS ensures the generation of new states. In the unstrained system, under the spin orbit interaction, these six states split into one four-fold and one two-fold degenerate points at $\Gamma$, as shown in Fig.\,\ref{qe_bands}(b). In this way, each of those three single-fold bands give rise to three two fold degeneracies at $\Gamma$ in the strained systems consistent with the results obtained from the DFT [Fig.\,\ref{qe_bands} (a) and (c)]. \\

\section{Summary} 

We studied CoSi system under biaxial strain using first-principles computations and  found that the degeneracies, causing multifold fermions at the time reversal symmetric points $\Gamma$ and $R$ without strain, are lifted due to the symmetry lowering. The appearance of equivalent Weyl points in the $k_{x}-k_{y}$ plane occurs in the strained systems. The distribution of these two-fold Weyl nodes are controlled by the direction and value of external strain. From our wannier-based surface state calculations, we found that the extended Fermi arcs connecting the bulk projections $\Bar{\Gamma}$ and $\Bar{M}$ are present on the (001) surface even in the strained systems. These connect Weyl points around $\Bar{\Gamma}$ and $\Bar{M}$ with opposite chirality. In contrast, we find that the (111) surface of the strained systems host small Fermi arcs near surface BZ center, unlike the unstrained system. We complement our first-principles results with a low energy effective model as well as symmetry analyses. We hope our findings motivate experimental explorations of strain-engineering of multifold fermions in the near future.\\

\section*{Acknowledgments}

We acknowledge Peizhe Tang, Tiantian Zhang, Aftab Alam and Dmitry Pshenay-Severin for useful discussions. AB thanks Prime Minister's Research Fellowship for financial support. AN acknowledges support from the start-up grant (SG/MHRD-19-0001) of the Indian Institute of Science and from DST-SERB (project number SRG/2020/000153).\\

\section{Appendix}

\subsection{Unit cell parameters and atomic positions}
TABLE \ref{table:unitcell} contains lattice parameters for the unit cell for different strain values upto second decimal place. Atomic positions for these structures are listed in TABLE \ref{table:atomic_coordinates}.

\begin{table*}[h!]
\caption{Unit cell parameters.}
\label{table:unitcell}
\begin{tabular}{|p{2.2cm}|p{1.8cm}|p{1.6cm}|p{2.1cm}|}
 \hline
 %\multicolumn{4}{|c|}{Unit cell information} \\
 \hline
 Strain values & $a=b$ ($\AA$) & $c$ ($\AA$)& $\alpha=\beta=\gamma$\\
 \hline
  $+2\%$  &4.52&4.36& $90^{\circ}$\\
 
  Unstrained &4.43& 4.43& $90^{\circ}$\\

  $-2\%$ &4.34& 4.49& $90^{\circ}$\\

 \hline
\end{tabular}
\end{table*}

\begin{table*}[h!]
\caption{Relaxed atomic positions (in crystal coordinates).} % title name of the table
\label{table:atomic_coordinates}
\centering % centering table
\begin{tabular}{l  c rrr} % creating 3 columns
\hline\hline % inserting double-line
 Strain-value &Atom &\multicolumn{3}{c}{Position of atoms}
\\ [0.5ex]
\hline % inserts single-line
% Entering 1st row
  &Co &0.857037815 & 0.643837745 & 0.355409707 \\[-1ex]
\raisebox{1.5ex}{+2\%} &Co
&0.642962185 & 0.356162255 & 0.855409707  \\[0ex]
 &Co
& 0.357037815 & 0.856162255 & 0.644590293  \\[0ex]
 &Co
& 0.142962185 & 0.143837745 & 0.144590293  \\[0ex]
 &Si
& 0.158026586 & 0.342493710 & 0.655768750  \\[0ex]
 &Si
& 0.341973414 & 0.657506290 & 0.155768750  \\[0ex]
 &Si
&  0.658026586& 0.157506290 & 0.344231250  \\[0ex]
 &Si
& 0.841973414 & 0.842493710 &0.8442312504  \\[2ex]
% Entering 2nd row
 &Co & 0.854900000& 0.645100000 & 0.354900000 \\[-1ex]
\raisebox{1.5ex}{unstrained} &Co
&0.645100000 & 0.354900000 & 0.854900000  \\[0ex]
 &Co
& 0.354900000 & 0.854900000 & 0.645100000  \\[0ex]
 &Co
& 0.145100000 & 0.145100000 & 0.145100000  \\[0ex]
 &Si
& 0.156804000 & 0.343196000 & 0.656804000  \\[0ex]
 &Si
& 0.343196000 & 0.656804000 & 0.156804000  \\[0ex]
 &Si
&  0.656804000& 0.156804000 & 0.343196000  \\[0ex]
 &Si
& 0.843196000 &0.843196000 &0.843196000  \\[2ex]

% Entering 3rd row
 &Co & 0.854104867& 0.645645840 & 0.355136373 \\[-1ex]
\raisebox{1.5ex}{-2\%} &Co
&0.645895133 & 0.354354160 & 0.855136373  \\[0ex]
 &Co
& 0.354104867 & 0.854354160 & 0.644863627  \\[0ex]
 &Co
& 0.145895133 & 0.145645840 & 0.144863627  \\[0ex]
 &Si
& 0.155635570 & 0.343747311 & 0.657324208  \\[0ex]
 &Si
& 0.344364430 & 0.656252689 & 0.157324208  \\[0ex]
 &Si
&  0.655635570& 0.156252689 &  0.342675792 \\[0ex]
 &Si
& 0.844364430 &0.843747311& 0.842675792 \\[2ex]
% [1ex] adds vertical space
\hline % inserts single-line
\end{tabular}
%\label{tab:PPer}
\end{table*}

\subsection{Surface bands for finite sized systems}
\begin{figure*}
\centering
\includegraphics[scale=0.25]{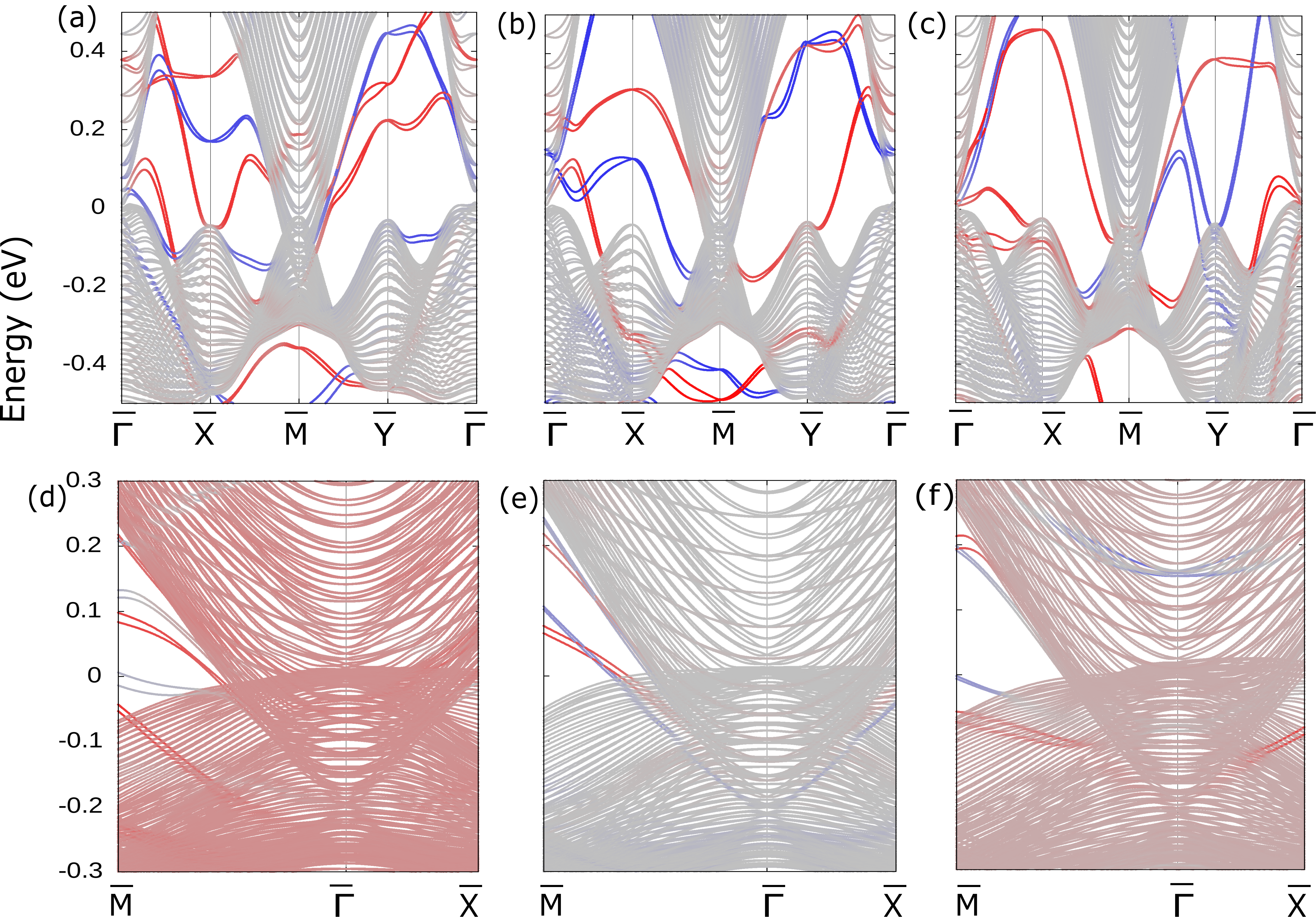}
  \caption{\textbf{(001) and (111) Surface bands for finite sized systems.} (a)-(c) \& (d)-(f) show surface bands for the 2\% tensile strained, unstrained and 2\% compressive strained systems for (001) surface of a slab with 20 unit cell layers \& (111) surface with 50 unit cell layers respectively. Grey coloured bands denote bulk bands, where as blue and red coloured bands denote the contribution coming from opposite surfaces of the slab.}
  \label{surface_001_111}
\end{figure*}

In addition to the surface bands for semi-infinite system in \ref{surface}, we have also calculated the bands for the finite sized slab. Fig.\,\ref{surface_001_111}
(a)-(c) and (d)-(f) present the bands for (001) and (111) surfaces for a finite slab of 20 and 50 unit cell layers respectively. The results obtained from this calculation are consistent with those obtained in Fig.\,\ref{fermi_arcs} and Fig.\,\ref{fermi_arcs_111}.\\

\subsection{Validity of the model Hamiltonian:} 

In order to prove the validity of the model Hamiltonian, here we present a comparison between the bands obtained from DFT calculation and the low energy model Hamiltonian.

\begin{figure*}
%\centering
\includegraphics[scale=0.25]{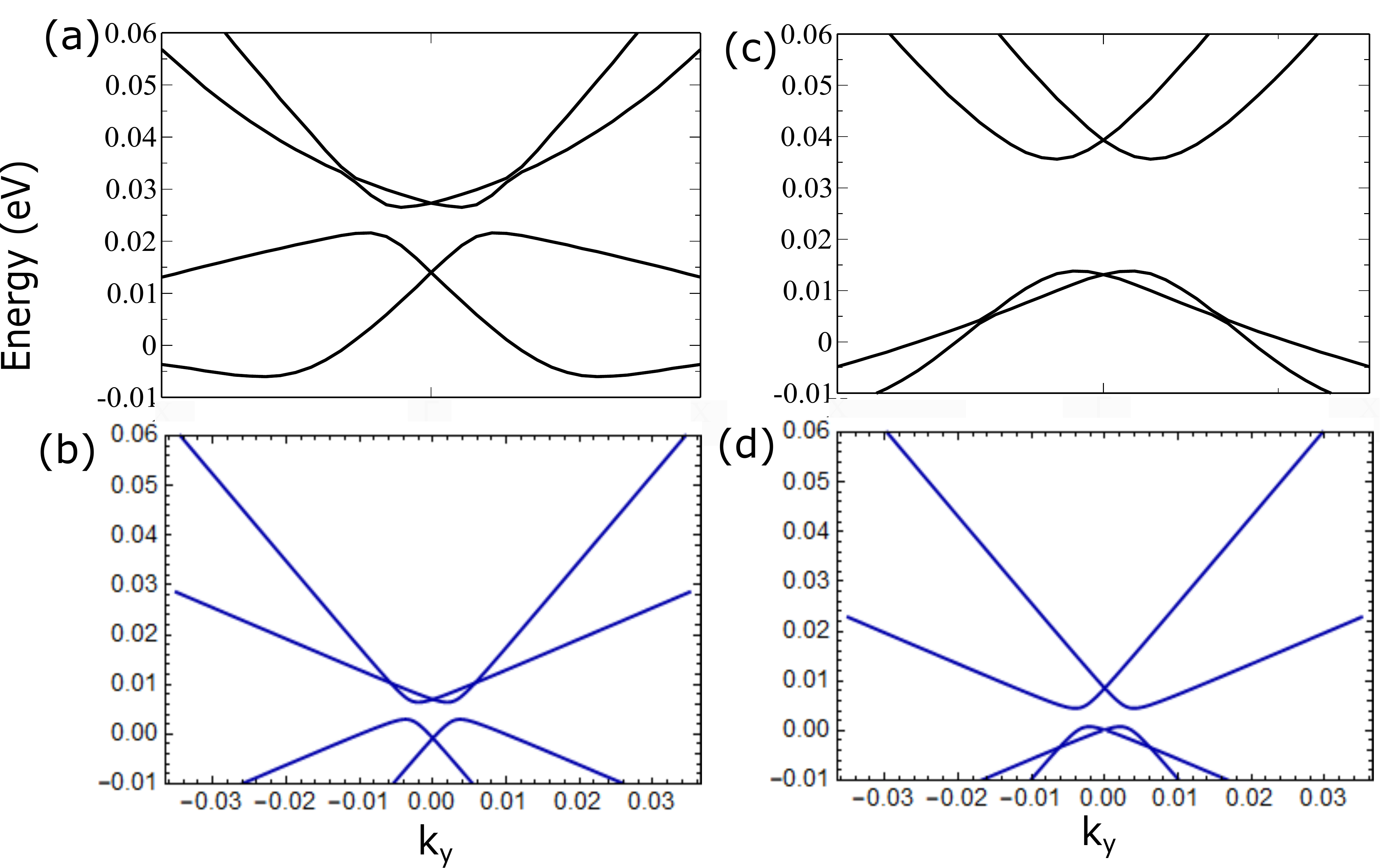}
\caption{\textbf{Comparison of DFT and model Hamiltonian band structures.} Top (marked with black) and bottom (marked with blue) panels show the bands obtained from DFT calculations and from the model Hamiltonian, respectively. (a) and (b) represent the band structures for the 2\% tensile strained system.  (c) and (d) represent the band structures for the 2\% compressive system. (a) and (c) (also, (b) and (d)) show that the bands obtained from DFT and model Hamiltonian match qualitatively near $\Gamma$.}
  \label{dft_model}
\end{figure*}

Fig.\,\ref{dft_model} shows a comparison between the bands along $k_y$ direction obtained from the model Hamiltonian to those obtained from DFT calculation for the 2\% tensile and compressive strained systems. We find that all the qualitative features from the DFT results are well reproduced by the model. In particular, the nature of the band crossings at and around $\Gamma$ match in both cases.

%\bibliography{references}


\end{document}